\documentclass[aps,prd,amsmath,amssymb,showpacs,11pt]{revtex4-1}
\usepackage{graphicx,color,dcolumn}
\usepackage{amsmath}
\usepackage{epsfig,epsf,psfrag}
\usepackage{bm}
\usepackage{amssymb}
\usepackage{amsfonts}
\usepackage{keyval,graphicx}
\usepackage{textcomp,wasysym}
\usepackage{epstopdf}
\begin{document}

\title{Disentangling the role of the $Y(4260)$ in $e^+e^-\to D^*\bar{D}^*$ and
$D_s^*\bar{D}_s^*$ via line shape studies }

\author{Si-Run Xue$^{1,3,}$\footnote{{\it E-mail address:} xuesr@ihep.ac.cn},
    Hao-Jie Jing$^{2,3,}$\footnote{{\it E-mail address:} jinghaojie@itp.ac.cn},
    Feng-Kun Guo$^{2,3,}$\footnote{{\it E-mail address:} fkguo@itp.ac.cn},
and Qiang Zhao$^{1,3,4,}$\footnote{{\it E-mail address:} zhaoq@ihep.ac.cn}}

\affiliation{$^1$Institute of High Energy Physics and Theoretical Physics Center
for Science Facilities,\\
         Chinese Academy of Sciences, Beijing 100049, China}

\affiliation{$^2$CAS Key Laboratory of Theoretical Physics, Institute of
Theoretical Physics, Chinese Academy of Sciences, Beijing 100190, China}

\affiliation{$^3$School of Physical Sciences, University of Chinese Academy of
Sciences, Beijing 100049, China}

\affiliation{$^4$Synergetic Innovation Center for Quantum Effects and
Applications (SICQEA), Hunan Normal University, Changsha 410081, China}

% ==============================================================================
\begin{abstract}

Whether the $Y(4260)$ can couple to open charm channels has been a crucial issue
for understanding its nature. The available experimental data suggest that the
cross section line shapes of exclusive processes in $e^+e^-$ annihilations have
nontrivial structures around the mass region of the $Y(4260)$.
As part of a series of studies of the $Y(4260)$ as mainly a
$\bar{D}D_1(2420)+c.c.$ molecular state, we show that the partial widths of the
$Y(4260)$ to the two-body open charm channels of $e^+e^-\to D^*\bar{D}^*$ and
$D_s^*\bar{D}_s^*$ are much smaller than that to $\bar{D}D^*\pi+c.c.$.  The line
shapes measured by the Belle Collaboration for these two channels can be well
described by the vector charmonium states $\psi(4040)$, $\psi(4160)$ and
$\psi(4415)$ together with the $Y(4260)$. It turns out that the interference of
the $Y(4260)$ with the other charmonia produces a dip around 4.22~GeV in the
$e^+e^-\to D^*\bar{D}^*$ cross section line shape. The data also show an
evidence for the strong coupling of the $Y(4260)$ to the $D\bar D_1(2420)$, in
line with the expectation in the hadronic molecular scenario for the $Y(4260)$.

\end{abstract}


%14.40.Pq  Heavy quarkonia
%14.40.Rt  Exotic mesons

\date{\today}
\maketitle

\section{Introduction}

The mysterious state $Y(4260)$ has attracted a lot of attention since its
observation in 2005 by the BaBar Collaboration~\cite{Aubert:2005rm}. Although
many different models were proposed as solutions in the literature, it is
unfortunate that not all of these scenarios have been systematically studied and
compared with the existing experimental data (see e.g. several recent
reviews~\cite{Chen:2016qju,Lebed:2016hpi,Esposito:2016noz,Guo:2017jvc} for
summaries of some theoretical interpretations proposed in the literature).
Following a series of recent studies by treating the $Y(4260)$ as mainly a
$\bar{D}D_1(2420)+c.c.$ hadronic molecule, we are motivated to examine as many
as possible exclusive processes where the $Y(4260)$ can contribute. Such
systematic studies with more experimental constraints would either support or
invalidate the picture of the $Y(4260)$ being a hadronic molecule of
$\bar{D}D_1(2420)+c.c.$ and should provide more insights into its intrinsic
structure.  Therefore, we investigate the cross section line shapes of the
$e^+e^-\to D^*\bar{D}^*$ and $D_s^*\bar{D}_s^*$ processes which cover the mass
region of the $Y(4260)$ and contain several established conventional charmonium
states, i.e., $\psi(4040)$, $\psi(4160)$ and $\psi(4415)$.

So far, it has been demonstrated that most of the puzzling observations in the
mass region of $Y(4260)$ in $e^+e^-$ annihilations can be accounted for in the
same framework self-consistently. For the strong $S$-wave interactions between
$\bar{D}$ and $D_1(2420)$ (the charge conjugation, $D\bar{D}_1(2420)$, is always
implicated in the calculations), the dynamically generated $Y(4260)$ should
contain a large molecular component of $\bar{D}D_1(2420)+c.c.$ as the
long-distance component of its wave function, while a small short-distance
component is always allowed. The consequence is that the $Y(4260)$ will
dominantly decay into $\bar{D}D^*\pi+c.c.$ via the decays of its constituent
hadrons~\cite{Wang:2013kra,Cleven:2013mka,Qin:2016spb}. Moreover, due to the
strong $S$-wave coupling to the nearby $\bar D D_1(2420)$ channel, the cross
section line shape for the $e^+e^-\to \bar{D}D^*\pi+c.c.$ process should not be
described by a Breit--Wigner parametrization. This is generally true for any
states that strongly couple to nearby thresholds via an $S$-wave interaction.
Namely, it is natural to expect a nontrivial cross section line shape for
$e^+e^-\to \bar{D}D^*\pi+c.c.$ around the mass of the $Y(4260)$. This phenomenon
has been investigated in detail in Refs.~\cite{Cleven:2013mka,Qin:2016spb} which
are closely correlated with the study of the nature of the charged charmonium
states $Z_c(3900)$. The experimental data, i.e., the cross sections for
$e^+e^-\to J/\psi\pi\pi$, $h_c\pi\pi$ and the invariant mass spectra as well as
angular distributions of $Y(4260)\to \bar{D}D^*\pi+c.c.$, which were also
motivated by the search for $Z_c(3900)$ and $Z_c(4020)$ at
BESIII~\cite{Ablikim:2013mio,Ablikim:2013wzq,Ablikim:2013xfr,Ablikim:2014dxl,Ablikim:2015tbp,Ablikim:2015tbp,Ablikim:2015vvn},
have provided important constraints on the molecular component of the $Y(4260)$.

One interesting question arising from the above mentioned analysis is whether
the $Y(4260)$ should have significantly large decay widths into other open charm
channels apart from $\bar{D}D^*\pi+c.c.$. Given that the total width of
$Y(4260)$ is dominated by the $\bar{D}D^*\pi+c.c.$ channel~\cite{Gao:2017sqa}, which has a partial
width of about 65~MeV in Ref.~\cite{Qin:2016spb}, while its decays into the
hidden charm channels, i.e.
$J/\psi\pi\pi$, $h_c\pi\pi$, and $\chi_{c0}\omega$, turn out to be relatively
small, the $Y(4260)$ decays into other open charm channels should also have
small widths in order to match the total width extracted in the combined
analysis of $e^+e^-\to J/\psi\pi\pi$, $h_c\pi\pi$, and $\bar{D}D^*\pi+c.c.$ In
this sense, to accommodate the experimental data for $e^+e^-\to D^*\bar{D}^*$
and $D_s^*\bar{D}_s^*$ in the same framework is a challenge for the molecular
picture, and should provide more information about its structure.

In this work, we analyze the cross section line shapes of the $e^+e^-\to
D^*\bar{D}^*$ and $D_s^*\bar{D}_s^*$ processes from threshold to about 4.6~GeV.
These two processes have been measured by the Belle Collaboration using the
initial state radiation~(ISR) in $e^+e^-$
annihilations~\cite{Abe:2006fj,Pakhlova:2010ek}. One can see that the cross
sections for $e^+e^-\to D^*\bar{D}^*$ have been measured with a high
precision~\cite{Abe:2006fj}, but there are still large uncertainties in the data
for $e^+e^-\to D_s^*\bar{D}_s^*$~\cite{Pakhlova:2010ek}. The former process has
been studied in~\cite{Du:2016qcr} which considers the $P$-wave coupled-channel
effects due to a pair of ground state charmed mesons and the $\psi(4040)$ but
not the $Y(4260)$.
In our analysis, in addition to the $Y(4260)$ which is included as a
$\bar{D}D_1(2420)$ hadronic molecule, we also include several conventional
vector charmonium states established in this mass region including the
$\psi(4040)$, the $\psi(4160)$ and the $\psi(4415)$.
We try to understand the behavior of the molecular state $Y(4260)$ in this
energy region and its interference with other charmonium states in the
description of the cross section line shapes. We note in advance that our focus
is mainly in the vicinity of the $Y(4260)$, i.e. around the threshold of
$\bar{D}D_1(2420)$. Although there are additional exotic candidates above the
$\bar{D}D_1(2420)$ threshold, such as the $Y(4360)$, to be neglected in this analysis, we
find that we can still draw a clear conclusion on the $Y(4260)$ contribution due
to the relatively isolated $\bar{D}D_1(2420)$ threshold.

In this paper, we first estimate the partial decay width of $Y(4260)\to
D^*\bar{D}^*$ in the molecular picture in Sec.~II, and then we study the cross
section line shapes of $e^+e^-\to D^*\bar{D}^*$ and $D_s^*\bar{D}_s^*$
considering the $Y(4260)$ and three charmonium states mentioned above in
Sec.~III. A brief summary will be given in Sec.~IV.

\section{The partial decay width of $Y(4260)\to D^*\bar{D}^*$}

\begin{figure}[htb]
\centering
\scalebox{0.6}{\includegraphics{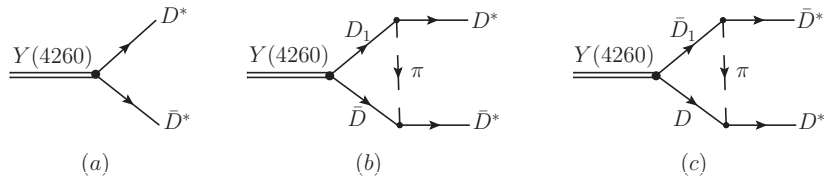}}
\caption{Feynman diagrams for the two-body decay $Y(4260)\to D^*\bar{D}^*$ in
our scenario.}
\label{fig-1}
\end{figure}

In our scenario, the $Y(4260)$ is treated as mainly an $S$-wave molecule of
$\bar{D}D_1(2420)+c.c.$ with a small mixture of a compact $c\bar{c}$
core~\cite{Qin:2016spb}. This treatment recognizes the HQSS breaking in the
production of $Y(4260)$ via $e^+e^-$ annihilations. Namely, its production in
$e^+e^-$ annihilations is mainly via the direct coupling to its compact
$c\bar{c}$ core which contains the $^3S_1(c\bar{c})$ configuration.
Then, the HQSS breaking allows the mixture of the $^3S_1(c\bar{c})$ core with
the long-distance component of $\bar{D}D_1(2420)+c.c.$ which can couple to
$^3D_1(c\bar{c})$ via an $S$-wave interaction. The wave function renormalization
will dress the nonvanishing $\gamma^*$--$^3S_1(c\bar{c})$ coupling and the
coupling of $Y(4260)$ to $\bar{D}D_1(2420)+c.c.$ as investigated in
Ref.~\cite{Qin:2016spb}. As a result of this scenario, it allows for the decay
of $Y(4260)\to D^*\bar{D}^*$ to occur not only via the dominant
$\bar{D}D_1(2420)+c.c.$ component but also through the direct coupling of
the $c\bar{c}$ core to $D^*\bar{D}^*$ as illustrated in Fig.~\ref{fig-1}.

In the framework of non-relativistic effective field theory (NREFT) the
Lagrangians for the coupling vertices in Fig.~\ref{fig-1} can be
written as~\cite{Wang:2013cya,Guo:2013nza,Cleven:2013mka,Qin:2016spb}
\begin{eqnarray}
  \label{eq:LY}
\mathcal{L}_{YD_1\bar{D}}&=&i\frac{y^{\text{eff}}}{\sqrt{2}}
(\bar{D}_a^\dagger Y^iD^{i\dagger}_{1a}-\bar D^{i\dagger}_{1a} Y^iD^\dagger_a)
+H.c. , \\
\label{eq:LD1}
\mathcal{L}_{D_1D^*\pi}&=&i\frac{h^\prime}{f_\pi}\Big[3D^i_{1a}
(\partial^i\partial^j\phi_{ab})D^{*\dagger j}_b-D^i_{1a}
(\partial^j\partial^j\phi_{ab})D^{*\dagger i}_b+3\bar D^i_{1a}
(\partial^i\partial^j\phi_{ba})\bar D^{*\dagger j}_{b}-\bar D^i_{1a}
(\partial^j\partial^j\phi_{ba})\bar D^{*\dagger i}_{b}\Big]\nonumber \\
&& +H.c.,\\
\label{eq:LDstar}
\mathcal{L}_{D^*D\pi}&=&g_{\pi}\big(D_a\partial^i\phi_{ab}D_b^{*i\dagger}+\bar{D}_a\partial^i\phi_{ba}\bar{D}_b^{*i\dagger}\big)+H.c. ,
\end{eqnarray}
where $f_\pi=132$ MeV and the effective coupling for $Y(4260)$ and
$\bar{D}D_1(2420)$ is $y^{\text{eff}}=(3.94\pm0.04)\text{GeV}^{-1/2}$ which has
been determined by the combined analysis of $e^+e-\to J/\psi\pi\pi$, $h_c\pi\pi$
and $\bar{D}D^*\pi+c.c.$~\cite{Cleven:2013mka,Qin:2016spb}; the effective
coupling constants $h^\prime$ and $g_{\pi}$ can be determined by the processes
of $D_1^0 \to D^{*+}\pi^-$ and $D^{*-} \to D^{0}\pi^-$, respectively. The direct
coupling for $Y(4260)\to D_{(s)}^*\bar{D}_{(s)}^*$ takes the same form as the
vector charmonium couplings to $D_{(s)}^*\bar{D}_{(s)}^*$ and will be given in
the next section.

The decay amplitude for the loop diagrams in Fig.~\ref{fig-1}~(b) and (c) can
then be expressed as
\begin{eqnarray}
\mathcal{M}^{\text{Loop}}_{Y(4260)\to D^*\bar{D}^*}&=&
\frac{3y^{\text{eff}} h^\prime g_\pi}{2\sqrt{2}f_\pi}\epsilon_Y^i
\epsilon_{D^*}^{j*}\epsilon_{\bar{D}^*}^{k*}
\int\frac{d^4l}{(2\pi)^4} \nonumber\\
&&\times\Bigg\{
\frac{3l^il^jl^k-\delta^{ij}l^k\vec{l}^{\,2}}{[(p_1+l)^2-m_{D_1}^2+i0^+][(p_2-l)^2-m_D^2+i0^+][l^2-m_\pi^2+i0^+]}\nonumber
\\
&&- \frac{3l^il^jl^k-\delta^{ik}l^j\vec{l}^{\,2}}
{[(p_2+l)^2-m_{D_1}^2+i0^+][(p_1-l)^2-m_D^2+i0^+][l^2-m_\pi^2+i0^+]} \Bigg\}
\nonumber
\\
&\equiv&\frac{3y^{\text{eff}} h^\prime g_\pi}{2\sqrt{2}f_\pi}\epsilon_Y^i
\epsilon_{D^*}^{j*}\epsilon_{\bar{D}^*}^{k*}
\left[3I^{ijk}-C^{ijk}-3I^{ijk}(p_1\leftrightarrow
p_2)+C^{ikj}(p_1\leftrightarrow p_2)\right]\nonumber \\
&=&\frac{3y^{\text{eff}} h^\prime g_\pi}{2\sqrt{2}f_\pi}
\epsilon_Y^i\epsilon_{D^*}^{j*}\epsilon_{\bar{D}^*}^{k*}
\left(6I^{ijk}-C^{ijk}-C^{ikj}\right),
\end{eqnarray}
where $p_1$, $p_2$ and $l$ are the four momenta of the $D^*$, $\bar{D}^*$ and
$\pi$, respectively. In the last step, we have used $p_1^2=p_2^2=m_{D^*}^2$ and
$p_2^i=-p_1^i$ in the center-of-mass (c.m.) frame. The factor of ${3}/{2}$
comes from the isospin symmetry and function $C^{ijk}$ and $I^{ijk}$ are defined
as follows:
\begin{eqnarray}
C^{ijk}&\equiv &\sum_{m=1}^3\delta^{ij}I^{kmm},\\
I^{ijk}&\equiv &\int\frac{d^4l}{(2\pi)^4}
\frac{l^il^jl^k}{[(p_1+l)^2-m_{D_1}^2+i0^+][(p_2-l)^2-m_D^2+i0^+]
[l^2-m_\pi^2+i0^+]}\,.
\end{eqnarray}

It is interesting to compare the transition of Fig.~\ref{fig-1} with the hidden
charm decay channels such as $Y(4260)\to
Z_c(3900)\pi$~\cite{Wang:2013cya,Qin:2016spb} and
$\chi_{c0}\omega$~\cite{Cleven:2016qbn}. Following the NREFT power counting
scheme of Refs.~\cite{Guo:2009wr,Guo:2010ak,Guo:2017jvc}, it can be seen that
the loop amplitude for the $Y(4260)\to Z_c(3900)\pi$ is ultraviolet (UV)
convergent and scales as $1/v$ with $v$ the typical non-relativistic velocity of
the intermediate charmed mesons. With $v\ll1$ the loop integral gets enhanced in
comparison with the tree diagram. The case for the
$Y(4260)\to\chi_{c0}\omega$ is similar. Such a power counting is because of the
$S$-wave couplings of both the initial and final heavy particles to the
intermediate charmed mesons.

For the $Y(4260)\to D^*\bar{D}^*$,  the velocity scaling is different. Near the
mass threshold of $\bar{D}D_1(2420)$ the internal charmed mesons carry the
typical velocity $v\sim (|m_Y-m_D-m_{D_1}|/\tilde{m})^{1/2}\simeq0.1$, where
$\tilde{m}\equiv (m_D+m_{D_1})/2$, and the velocity of the final $D^*$ is
$v_f\simeq0.35$ in the $Y(4260)$ rest frame. For the velocity scaling, we may
count $v_f\sim v$. Then the loop integral measure scales as $v^5$, and all
propagators scale as $v^{-2}$. As a result, the triangle loop amplitude scales
as $v^5 v^{-6} v^3=v^2$, where the factor of $v^3$ comes from the vertices,
which is significantly suppressed in respect of the contact interaction.
Because of the $D$- and $P$-wave pionic couplings given by Eqs.~\eqref{eq:LD1}
and \eqref{eq:LDstar} , respectively, the loop decay amplitude can be split into
$P$-wave and $F$-wave parts as
\begin{eqnarray}
  I^{ijk} = \vec
  p_1^{\,2} \left( p_1^i\delta^{jk}+p_1^j\delta^{ik}+p_1^k\delta^{ij}\right) I_P
  + \left[p_1^ip_1^jp_1^k - \frac{1}{5} \vec p_1^{\,2} \left(
  p_1^i\delta^{jk}+p_1^j\delta^{ik}+p_1^k\delta^{ij}\right) \right] I_F .
  \label{eq:Ijik_PF}
\end{eqnarray}
The first term contributes to the decay into the $D^*\bar D^*$ in a $P$-wave,
while the second contributes to that in a $F$-wave. While the $F$-wave part is
UV convergent, the $P$-wave part diverges and needs to be regularized and renormalized. The UV divergence can be absorbed by introducing a
counterterm. However, the tree-level term of Fig.~\ref{fig-1} (a) cannot serve
as the counterterm for the loop amplitude of Fig.~\ref{fig-1} (b) and (c) since
diagram (a) is introduced to incorporate the $^3S_1(c\bar{c})$ coupling to
$D^*\bar{D}^*$ while in diagrams (b) and (c) the $S$-wave $\bar{D}D_1(2420)$,
which leads to the transitions to $D^*\bar{D}^*$, couples to the
$^3D_1(c\bar{c})$ in the heavy quark limit~\cite{Li:2013yka}. This means that
the UV divergence here needs to be absorbed into a different counterterm. Here
we will regularize the UV divergence practically using a form factor with a
cutoff, see below.
The cutoff will be treated as a free parameter, which effectively takes the
place of the counterterm at a given scale.

In order to regularize the UV contributions in the loop integral, we introduce a
monopole form factor for each propagator to take into account the off-shell
effects in the loop integral:
\begin{equation}
\mathcal{F}(\Lambda_i^2,m_i^2,l_i^2)=\frac{\Lambda_i^2-m_i^2}{\Lambda_i^2-l_i^2} \ ,
\end{equation}
where $\Lambda_1\equiv m_{D_1}+\alpha \Lambda_{\text{QCD}}$ and $\Lambda_2\equiv
m_{D}+\alpha \Lambda_{\text{QCD}}$, with $\Lambda_{\text{QCD}}=220$~MeV and
$\alpha$ a parameter of order unity, are defined for the heavy charmed mesons.
For the light pion exchange the cut-off $\Lambda_\pi$ is within a range of $0.5\sim 1$~GeV as usually adopted.
Then the loop amplitude $I^{ijk}$ can be expressed as:
\begin{eqnarray}
I^{ijk}&=&\int\frac{d^Dl}{(2\pi)^D}\frac{l^il^jl^k\mathcal{F}(\Lambda_1^2,m_{D_1}^2,(p_1+l)^2)\mathcal{F}(\Lambda_2^2,m_D^2,(p_2-l)^2)\mathcal{F}(\Lambda_\pi^2,m_\pi^2,l^2)}
          {[(p_1+l)^2-m_{D_1}^2+i0^+][(p_2-l)^2-m_D^2+i0^+][l^2-m_\pi^2+i0^+]}\nonumber \\
      &=&\frac{i}{16\pi^2}[p_1^ip_1^jp_1^kA^\prime+(-p_1^i\delta^{jk}-p_1^j\delta^{ik}-p_1^k\delta^{ij})B^\prime],
\end{eqnarray}
with
\begin{eqnarray}
A^\prime&\equiv &A(P^2,p_1^2,p_2^2,m_{D_1}^2,m_D^2,m_\pi^2)-A(P^2,p_1^2,p_2^2,\Lambda_1^2,m_D^2,m_\pi^2)-A(P^2,p_1^2,p_2^2,m_{D_1}^2,\Lambda_2^2,m_\pi^2)\nonumber \\
         &&-A(P^2,p_1^2,p_2^2,m_{D_1}^2,m_D^2,\Lambda_\pi^2)+A(P^2,p_1^2,p_2^2,\Lambda_1^2,\Lambda_2^2,m_\pi^2)+A(P^2,p_1^2,p_2^2,\Lambda_1^2,m_D^2,\Lambda_\pi^2)\nonumber\\
         &&+A(P^2,p_1^2,p_2^2,m_{D_1}^2,\Lambda_2^2,\Lambda_\pi^2)-A(P^2,p_1^2,p_2^2,\Lambda_1^2,\Lambda_2^2,\Lambda_\pi^2),\\
B^\prime&\equiv &B(P^2,p_1^2,p_2^2,m_{D_1}^2,m_D^2,m_\pi^2)-B(P^2,p_1^2,p_2^2,\Lambda_1^2,m_D^2,m_\pi^2)-B(P^2,p_1^2,p_2^2,m_{D_1}^2,\Lambda_2^2,m_\pi^2)\nonumber \\
         &&-B(P^2,p_1^2,p_2^2,m_{D_1}^2,m_D^2,\Lambda_\pi^2)+B(P^2,p_1^2,p_2^2,\Lambda_1^2,\Lambda_2^2,m_\pi^2)+B(P^2,p_1^2,p_2^2,\Lambda_1^2,m_D^2,\Lambda_\pi^2)\nonumber\\
         &&+B(P^2,p_1^2,p_2^2,m_{D_1}^2,\Lambda_2^2,\Lambda_\pi^2)-B(P^2,p_1^2,p_2^2,\Lambda_1^2,\Lambda_2^2,\Lambda_\pi^2) \ .
\end{eqnarray}
Here the functions $A$ and $B$ are defined as
\begin{eqnarray}
A(P^2,p_1^2,p_2^2,m_{D_1}^2,m_D^2,m_\pi^2)&=&\int_0^1dx\int_0^1\frac{y^4}{\Delta}dy,\\
B(P^2,p_1^2,p_2^2,m_{D_1}^2,m_D^2,m_\pi^2)&=&\int_0^1dx\int_0^1\frac{1}{2}y^2\ln\Delta dy,
\end{eqnarray}
where $P\equiv p_1+p_2$ is the initial momentum, $x$ and $y$ are the Feynman
parameters, and $\Delta=y^2p_x^2+(1-y)m_\pi^2-y\,\Delta m_x^2$ with $p_1 =
xp_1-(1-x)p_2$ and $\Delta m_x^2 = x(p_1^2-m_{D_1}^2)+(1-x)(p_2^2-m_{D}^2)$.
In the numerical calculation, we replace $m_{D_1}$ by $m_{D_1}-i\Gamma_{D_1}/2$
with $\Gamma_{D_1}$ the constant width of the $D_1(2420)$.

\begin{figure}[tb] \centering
\scalebox{1.}{\includegraphics{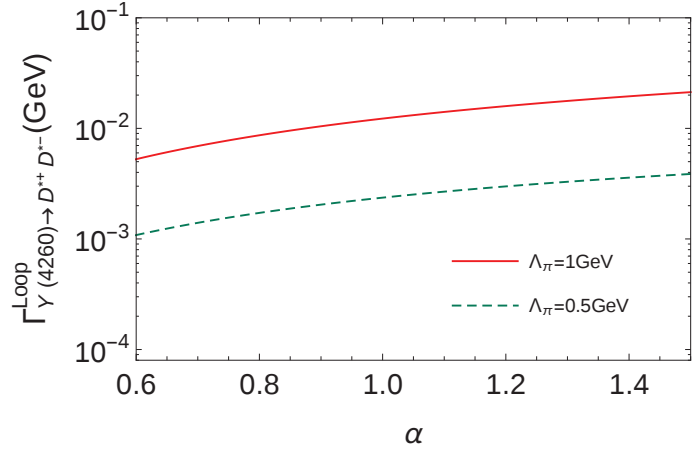}}
\caption{The cutoff-dependence of the partial decay width of $Y(4260)\to
D^{*+}D^{*-}$ from the one-pion exchange diagrams in the molecular scenario.
Here the results with two typical $\Lambda_\pi$ values are shown. }
\label{fig-2}
\end{figure}
Due to the UV divergence in the $P$-wave part of the loop amplitude, it is
impossible to make a definite prediction on the two-body decay partial width of
the $Y(4260)$ into a pair of vector charm mesons by simply calculating the loop
diagrams. The best we can do is to estimate the values by varying the cutoffs in
the form factors within natural ranges. Thus, in Fig.~\ref{fig-2} we show the
dependence of $\Gamma^{\text{Loop}}_{Y(4260)\to D^{*+}D^{*-}}$ on $\alpha$ with
two typical values for $\Lambda_\pi$. The result ranges from 1~MeV to about
20~MeV in the figure, and for $\alpha=1$ it takes a value of 2.4~MeV and
12.3~MeV for $\Lambda_\pi=0.5$~GeV and 1~GeV, respectively. These values are significantly smaller than the partial width for $Y(4260)\to \bar{D}D^*\pi+c.c.$

One intriguing feature of the $Y(4260)$ is that it does not show up as a peak in
the exclusive two-body open-charm cross sections.  In order to clarify the role
played by the $Y(4260)$ in $e^+e^-\to D^{*+}D^{*-}$, we will investigate the
cross section line shape of this process in the next section taking into account
contributions from the nearby charmonium states. The idea is to investigate
whether the cross section line shapes could provide more stringent constraint on
$Y(4260)$ or not. A combined investigation of the cross section line shape of
$e^+e^-\to D_s^{*+}D_s^{*-}$ will also be presented.

\section{The line shapes of the cross sections of $e^+e^-\to
D_{(s)}^*\bar{D}_{(s)}^*$}
\label{sec-3}

\begin{figure}[tb]
\centering
\scalebox{0.8}{\includegraphics{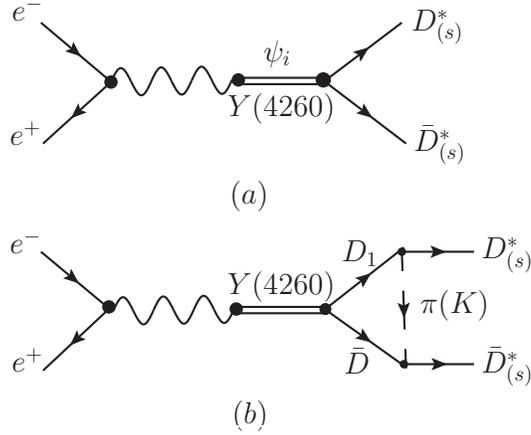}}
\caption{Feynman diagrams for $e^+e^- \to D_{(s)}^*\bar{D}_{(s)}^*$ via (a) intermediate charmonium states $\psi_i$, and (b) $Y(4260)$ as a $\bar{D}D_1(2420)+c.c.$ hadronic molecule state.} \label{fig-3}
\end{figure}

In order to investigate the line shape of $e^+e^-\to D_{(s)}^*\bar{D}_{(s)}^*$
in the vicinity of $Y(4260)$, the contributions from the nearby charmonium
states, $\psi(4040)$, $\psi(4160)$ and $\psi(4415)$, which are normally
considered as the $3S$, $2D$ and $4S$ charmonium states, respectively, should be
included.
For convenience, we use $\psi_1$, $\psi_2$ and $\psi_3$ to denote $\psi(4040)$,
$\psi(4160)$ and $\psi(4415)$, respectively. The processes of $e^+e^-\to
D_{(s)}^*\bar{D}_{(s)}^*$ are depicted in Fig.~\ref{fig-3} where the tree-level
diagram represents the charmonium transitions and the loop diagram illustrates
the $Y(4260)$ contribution via its molecular component. As mentioned earlier,
the tree diagram also contains the contribution from the short-distance core of
the $Y(4260)$.

The effective Lagrangian for the vector charmonium couplings to the virtual
photon is described by the vector meson dominance (VMD) model:
\begin{eqnarray}\label{vmd-coup}
\mathcal{L}_{V\gamma}&=&\frac{e m^2_V}{f_V}V_\mu A^\mu \ ,
\end{eqnarray}
while the strong couplings for $\psi_i$ ($i=1, \ 2, \ 3$) to the
$D_{(s)}^*\bar{D}_{(s)}^*$ meson pairs are as
follows~\cite{Margaryan:2013tta,Guo:2013nza}:
\begin{eqnarray}
\mathcal{L}_{\psi_SD_{(s)}^*\bar{D}_{(s)}^*}&=&
ig_{\psi_SD^*_{(s)}\bar{D}^*_{(s)}}\psi_S^{k}\big{[}(\delta^{km}\delta^{ln}-
\delta^{kn}\delta^{lm}-\delta^{kl}\delta^{mn})(\partial^mD_{(s)}^{{*\dag}l}
\bar{D}_{(s)}^{*\dag n}-{D}_{(s)}^{*\dag l}\partial^m\bar{D}_{(s)}^{*\dag n})
\big{]}+H.c. , \nonumber\\
\mathcal{L}_{\psi_DD_{(s)}^*\bar{D}_{(s)}^*}&=&
ig_{\psi_DD^*_{(s)}\bar{D}^*_{(s)}}\psi_D^{k}\big{[}(4\delta^{km}\delta^{ln}-
\delta^{kn}\delta^{lm}-\delta^{kl}\delta^{mn})
(\partial^mD_{(s)}^{{*\dag}l}\bar{D}_{(s)}^{*\dag n}-{D}_{(s)}^{*\dag l}
\partial^m\bar{D}_{(s)}^{*\dag n})\big{]}+H.c. , \nonumber\\
\end{eqnarray}
where the coupling constants $g_{\psi_iD^*_{(s)}\bar{D}^*_{(s)}}$ will be
determined by fitting the cross section line shapes. Note that $\psi_S$ and
$\psi_D$ denote the $S$ and $D$ wave $c\bar{c}$ states of $J^{PC}=1^{--}$ of
which the couplings to $D_{(s)}^*\bar{D}_{(s)}^*$ are different, and the above
forms are obtained assuming heavy quark spin symmetry.

For the $e^+e^- \to D^*\bar{D}^*$, we consider all of the three conventional
charmonium states mentioned above and the $Y(4260)$, while for the $e^+e^- \to
D_{s}^*\bar{D}_{s}^*$ we only include $\psi_2$ ($\psi(4160)$) and $\psi_3$
($\psi(4415)$) as the contributing charmonium states since
$\psi_1$ ($\psi(4040)$) is far below the threshold of $D_{s}^*\bar{D}_{s}^*$.
The transition amplitudes for $e^+e^- \to D_{(s)}^*\bar{D}_{(s)}^*$ can then be
expressed as
\begin{eqnarray}\label{trans-amp}
&& \mathcal{M}_{e^+e^- \to D^*\bar{D}^*} \nonumber\\
&=&
\bar{v}(q_2)\gamma^iu(q_1)\Bigg{\{}(2\delta^{ij}p_1^k+2\delta^{ik}p_1^j-
2\delta^{jk}p_1^i) \Bigg[\frac{-e^2g_{\psi_1D^*\bar{D}^*}m^2_{\psi_1}
\exp(-2|p_f^2-p_{10}^2|/\beta^2+i\theta_1)}
{f_{\psi_1}E_{\rm cm}^2(E_{\rm cm}^2-m_{\psi_1}^2+im_{\psi_1}\Gamma_{\psi_1})}\nonumber
\\
&&+\frac{-e^2g_{\psi_3D^*\bar{D}^*}m^2_{\psi_3}\exp(-2|p_f^2-p_{30}^2|/\beta^2+i\theta_3)}{f_{\psi_3}E_{\rm cm}^2(E_{\rm cm}^2-m_{\psi_3}^2+im_{\psi_3}\Gamma_{\psi_3})}
+\frac{-e^2g^{\text{eff}}_{YD^*\bar{D}^*}m_{Y}\exp(-2|p_f^2-p_{Y0}^2|/\beta^2)}
{f^{\text{eff}}_{Y}E_{\rm cm}^2\mathcal{D}_Y(E_{\rm cm})}\Bigg]\nonumber \\
&&+(2\delta^{ij}p_1^k+2\delta^{ik}p_1^j-8\delta^{jk}p_1^i)
\frac{-e^2g_{\psi_2D^*\bar{D}^*}m^2_{\psi_2}\exp(-2|p_f^2-p_{20}^2|/\beta^2+i\theta_2)}{f_{\psi_2}E_{\rm cm}^2(E_{\rm cm}^2-m_{\psi_2}^2+im_{\psi_2}\Gamma_{\psi_2})}\nonumber \\
&&+\frac{-3e^2y^{\text{eff}} h^\prime g_\pi m_{Y}} {2\sqrt{2}f_\pi
f_Y^{\text{eff}}E_{\rm cm}^2\mathcal{D}_Y(E_{\rm cm})}
\left[6I^{ijk}(\alpha,\Lambda_\pi)-
C^{ijk}(\alpha,\Lambda_\pi)-C^{ikj}(\alpha,\Lambda_\pi)\right]\Bigg{\}}
\epsilon_{D^*}^{j*}\epsilon_{\bar{D}^*}^{k*},
\end{eqnarray}
\begin{eqnarray}\label{trans-amp-Ds}
&& \mathcal{M}_{e^+e^- \to
D_s^*\bar{D}_s^*} \nonumber\\
&=&\bar{v}(q_2)\gamma^iu(q_1)\Bigg{\{}(2\delta^{ij}p_1^k+2\delta^{ik}p_1^j-
8\delta^{jk}p_1^i) \frac{-e^2g_{\psi_2D_s^*\bar{D}_s^*}m^2_{\psi_2}
\exp(-2|p_f^2-p_{20}^2|/\beta^{2}+i\theta_4)}{f_{\psi_2}E_{\rm cm}^2(E_{\rm
cm}^2-m_{\psi_2}^2+im_{\psi_2}\Gamma_{\psi_2})}\nonumber \\
&&+(2\delta^{ij}p_1^k+2\delta^{ik}p_1^j-2\delta^{jk}p_1^i)
\Bigg[\frac{-e^2g_{\psi_3D_s^*\bar{D}_s^*}m^2_{\psi_3}
\exp(-2|p_f^2-p_{30}^2|/\beta^{2}+i\theta_5)}{f_{\psi_3}
E_{\rm cm}^2(E_{\rm cm}^2-m_{\psi_3}^2+im_{\psi_3}\Gamma_{\psi_3})}\nonumber\\
&&+\frac{-e^2g^{\text{eff}}_{YD_s^*\bar{D}_s^*}m_{Y}\exp(-2|p_f^2-p_{Y0}^2|/\beta^2)}
{f^{\text{eff}}_{Y}E_{\rm cm}^2\mathcal{D}_Y(E_{\rm cm})}\Bigg]\nonumber \\
&&+\frac{-2e^2y^{\text{eff}} h^\prime g_\pi m_Y}{\sqrt{2}f_\pi f_Y^{\text{eff}}
E_{\rm cm}^2\mathcal{D}_Y(E_{\rm cm})}
\left[6I^{ijk}(\alpha,\Lambda_K)-C^{ijk}(\alpha,\Lambda_K)-
C^{ikj}(\alpha,\Lambda_K)\right]\Bigg{\}}\epsilon_{D_s^*}^{j*}\epsilon_{\bar{D}_s^*}^{k*},
\end{eqnarray}
where $q_1$ and $q_2$ are the incoming four-momenta of the electron and
positron, respectively, $\vec{p}_1$ is the outgoing momentum of $D_{(s)}^*$ in
the c.m. frame, $p_f=|\vec{p}_1|=\sqrt{E_{\rm cm}^2- 4m_{D_{(s)}^{*2}}}/2 $ is
the magnitude of the c.m. momentum of the final state, and $p_{i0} =
\sqrt{m_{\psi_i}^2- 4m_{D_{(s)}^{*2}}}/2$ is the magnitude of the momentum when
the c.m. energy is fixed to the intermediate charmonium mass.  Note that the pion and kaon propagators are implicated for the exchanged light mesons in the loop functions in Eqs.~(\ref{trans-amp}) and (\ref{trans-amp-Ds}), respectively. The Gaussian
form factor suppresses the resonance contributions when they become far
off-shell, and the parameter $\beta$ controls the suppression. As a
reasonable assumption to reduce the number of parameters, we assume that these
two processes share the same value for $\beta$ which means that the strong
couplings for $\psi_i$ to $D^*\bar{D}^*$ and $D_s^*\bar{D}_s^*$ have the same
suppression behavior when the resonances become off-shell. The $\psi_i$ states
and the $Y(4260)$ can interfere through many possible intermediate hadron loops
which can introduce energy-dependent complex phases. In order to parameterize
such effects, we also introduce a few constant phases, denoted by $\theta_i$.
The constant phase assumption is reasonable as long as the thresholds of the
intermediate hadrons are far away. Based on the above argument and taking into account the SU(3) flavor symmetry,
we let $\theta_4=\theta_2$ and $\theta_5=\theta_3$ to reduce two more
parameters.

In Eq.~(\ref{trans-amp}) $y^{\text{eff}}/[f_Y^{\text{eff}}\mathcal{D}_Y(E_{\rm
cm})]$ is the product of the bare coupling $y/f_Y$ and the $Y(4260)$ propagator
defined in the molecular picture~\cite{Qin:2016spb} which has the following
expression:
\begin{eqnarray}
\frac{y^{\text{eff}}}{f_Y^{\text{eff}}\mathcal{D}_Y(E_{\rm cm})}\equiv  \frac{Z
y}{2 f_Y [E_{\rm cm}-m_Y-Z\widetilde{\Sigma}_1(E_{\rm cm})+i\Gamma^{\text{non}-\bar{D}D_1}/2]} ,
\end{eqnarray}
where the subtracted self-energy $\widetilde{\Sigma}_1(E_{\rm cm}) =
\Sigma_1(E_{\rm cm}) - {\rm Re}\Sigma_1(m_Y) - (E-m_Y) \partial\Sigma_1(m_Y)/\partial E_{\rm
cm}$~\cite{Cleven:2011gp} with $\Sigma_1(E_{\rm cm})$ the $Y(4260)$ self-energy
due to the $\bar D D_1(2420)$ loop. In the $\overline{\rm MS}$ subtraction scheme, the self-energy is given by
$\Sigma_1(E_{\rm cm})=\mu/(8\pi)\, \sqrt{2\mu(E_{\rm cm}-m_D-m_{D_1})
+i\mu\Gamma_{D_1} }$~\cite{Qin:2016spb}. We use $m_Y=(4.217\pm0.002)$~GeV and
$\Gamma^{\text{non}-\bar{D}D_1}=(0.056\pm0.003)$~GeV are determined in the
combined analysis of $e^+e^-\to J/\psi\pi\pi$ and
$h_c\pi\pi$~\cite{Cleven:2013mka}, and the wave function renormalization
constant $Z\simeq0.13$ is determined in Ref.~\cite{Qin:2016spb}.
The values of $m_{\psi_i}$, $\Gamma_{\psi_i \to
e^+e^-}$ and $\Gamma_{\psi_i}$ ($i=1,2,3$) are taken from those given by the
Particle Data Group (PDG)~\cite{Olive:2016xmw}, which are listed in
Table~\ref{list-1}. The leptonic decay coupling constants of the charmonium
states defined by the VMD model in Eq.~(\ref{vmd-coup}) can thus be determined.

To further reduce the number of parameters we assume the SU(3) flavor symmetry
for the strong couplings of the same charmonium states to $D^*\bar{D}^*$ and
 $D_s^*\bar{D}_s^*$ so that they take the same value, i.e.
$g_{\psi_2D^*\bar{D}^*}=g_{\psi_2D_s^*\bar{D}_s^*}$ and $g_{\psi_3
D^*\bar{D}^*}=g_{\psi_3 D_s^*\bar{D}_s^*}$. In total there are 11
parameters to be fitted from the cross section data: four cutoff parameters
($\alpha$, $\Lambda_\pi$, $\Lambda_K$ and $\beta$), four coupling constants
($g_{\psi_i D^*\bar D^*}$ and $g_{Y D^*\bar D^*}^{\rm eff}$ which is the
coupling of the short-distance core of the $Y(4260)$ to the $D^*\bar D^*$), and
three phases. We note in advance that due to the lack of precise experimental
measurements for the $e^+e^- \to D_{s}^*\bar{D}_{s}^*$ some of the parameters
cannot be well constrained in the numerical fitting, and we anticipate that the
main contributions to the $\chi^2$ value will be from the $D^*\bar{D}^*$
channel.

\begin{table}[tb]
\caption{The masses, total widths and leptonic partial widths adopted for the charmonium states from PDG~\cite{Olive:2016xmw}. }\label{list-1}
\center
\begin{tabular}{|c|c|c|c|}
  \hline
  \hline
   & $\psi(4040)$ & $\psi(4160)$ & $\psi(4415)$ \\
  \hline
  $m_{\psi}$(MeV) & $4039\pm1$ & $4191\pm5$ & $4421\pm4$ \\
  \hline
  $\Gamma_{\psi}$(MeV)  & $80\pm10$ & $70\pm10$ & $62\pm20$\\
  \hline
  $\Gamma_{e^+e^-}$(keV) & $0.86\pm0.07$ & $0.48\pm0.22$ & $0.58\pm0.07$ \\
  \hline
  \hline
\end{tabular}
\end{table}

\begin{table}[tb]
\caption{Parameters determined by fitting to the Belle experimental
data~\cite{Abe:2006fj,Pakhlova:2010ek}.}\label{parameter-list}
\begin{tabular}{|c|c|}
  \hline
  \hline
  Parameters & Fitted values \\
  \hline
  $\alpha$ & $(1.01\pm0.12)$ \\
\hline
  $\beta$ & $(3.58\pm2.34)$ GeV \\
  \hline
  $\Lambda_\pi$ & $(409.1\pm23.9)$ MeV \\
\hline
 $\Lambda_K$ & $(544.7\pm71.9)$ MeV \\
\hline
  $\theta_{1}$ & $132.97^\circ\pm19.20^\circ$ \\
  \hline
  $\theta_{2,4}$ & $229.06^\circ\pm106.42^\circ$ \\
  \hline
 $\theta_{3,5}$ & $284.32^\circ\pm58.05^\circ$ \\
 \hline
  $g_{\psi_1D^*\bar{D}^*}$ & $(1.96\pm0.12)\ \text{GeV}^{-3/2}$ \\
  \hline
  $g_{\psi_2D_{(s)}^*\bar{D}_{(s)}^*}$ & $(0.11\pm0.13)\ \text{GeV}^{-3/2}$ \\
  \hline
  $g_{\psi_3D_{(s)}^*\bar{D}_{(s)}^*}$ & $(0.18\pm0.06)\ \text{GeV}^{-3/2}$ \\
  \hline
  $g^{\text{eff}}_{YD_{(s)}^*\bar{D}_{(s)}^*}$ & $(0.32\pm0.12)\ \text{GeV}^{-3/2}$ \\
  \hline\hline
  $\chi^2/$d.o.f & 1.53 \\
  \hline \hline
\end{tabular}
\end{table}
In Table~\ref{parameter-list} the values of the fitted parameters are listed.
The value of $\beta$, which bears a large uncertainty, is consistent with the
reasonable order of 1~GeV.
The cutoff parameter $\alpha$ is consistent with ${\cal O}(1)$. The cutoff
$\Lambda_\pi$ can be better constrained in $e^+e^-\to D^*\bar{D}^*$ than
$\Lambda_K$ in $e^+e^-\to D^*_s\bar{D}^*_s$, again due to the poor data quality
of the latter.
With the fitted $\alpha$ and $\Lambda_\pi$ values, the contribution from the
$\bar D D_1(2420)$ intermediate states to the partial decay width for
$Y(4260)\to D^*\bar{D}^*$ is given by $\Gamma^{\text{Loop}}_{Y(4260)\to
D^{*+}D^{*-}}=(1.27\pm0.68)$ MeV, while the contribution from the short-distance
$^3S_1$ $c\bar c$ core is much larger as listed in Table~\ref{widths-list}.
With the fitted couplings for the charmonium states to $D^*\bar{D}^*$ and
$D_s^*\bar{D}_s^*$, we can also obtain their corresponding partial decay widths
which are also listed in Table~\ref{widths-list}.

\begin{table}[tb]
\caption{The partial decay widths of $Y(4260)$ and $\psi_i$ to $D_{(s)}^*\bar{D}_{(s)}^*$ extracted from this analysis.  }\label{widths-list}
\center
\begin{tabular}{|c|c|c|c|c|}
  \hline
  \hline
  Widths & $Y(4260)$ & $\psi(4040)$ & $\psi(4160)$ & $\psi(4415)$ \\
   \hline
  $\Gamma_{YD^*\bar{D}^*}^{\text{Tree}}$ (MeV) & $9.50\pm7.18$ & - & - & -\\
   \hline
  $\Gamma_{YD^*\bar{D}^*}^{\text{Loop}}$ (MeV) & $1.27\pm0.68$ & - & - & -\\
  \hline
  $\Gamma_{D^*\bar{D}^*}$ (MeV) &  $10.77\pm7.86$ & $10.22\pm1.47$ & $4.74\pm11.73$ & $8.52\pm5.93$ \\
  \hline
  $\Gamma_{D_s^*\bar{D}_s^*}$ (MeV) & - & - & - & $3.36\pm2.34$\\
  \hline
  \hline
\end{tabular}
\end{table}

\begin{figure}[tb]
\centering
\scalebox{1}{\includegraphics{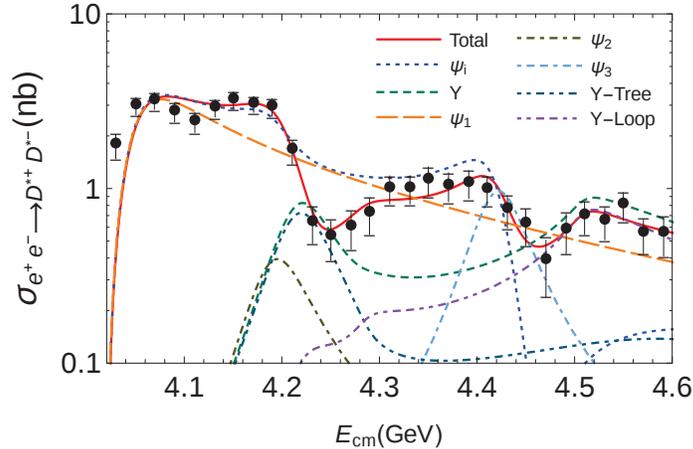}}\hspace{0.6cm}
\caption{The fitting results for the cross section of $e^+e^- \to D^{*+}D^{*-}$.
The overall cross section is denoted by the solid line. The exclusive
contributions from single states are also presented, i.e. $\psi(4040)$
(long-dashed), $\psi(4160)$ (dot-dashed), $\psi(4415)$ (dot-dashed-dashed), and
$Y(4260)$ (dashed). The sum of the contributions from all $\psi_i$
states is denoted by the dotted line. The data are from Ref.~\cite{Abe:2006fj}.}
\label{fig-4}
\end{figure}

With the fitted parameters in Table~\ref{parameter-list}, we find that the cross
section of $e^+e^-\to D^*\bar{D}^*$ can be well described. The line shape from
the best fit is plotted in Fig.~\ref{fig-4} and compared with the Belle
data~\cite{Abe:2006fj}. An apparent feature is that  the threshold enhancement
in the measured $e^+e^-\to D^*\bar{D}^*$ line shape can be largely accounted for
by the $\psi(4040)$ while the contributions from $Y(4260)$ and other charmonium
states are rather small below 4.2~GeV.
The bump between 4.1 and 4.2~GeV can be described well by the contributions from
$\psi(4040)$ and $\psi(4160)$ and their interference. Notice that the relative
phase between these two states leads to destructive interference in the energy
regions of below $\psi(4040)$ or above $\psi(4160)$ while in the region between
their masses the interference is constructive. As a result, the rise of the
cross section in the near threshold region is enhanced, although the
$\psi(4040)$ coupling to $D^*\bar{D}^*$ is in a $P$ wave.

Comparing the dotted curve (denoted by ``$\psi_i$'' in the figure), which is the
sum of the contributions from all considered conventional charmonium states,
with the solid curve, which is the best fit result, or with the experimental
data, one sees that the dip around 4.22~GeV in the data comes from a destructive
interference between the charmonia and the $Y(4260)$. The good description of
the special shape around 4.3~GeV originates from the strong coupling of the
$Y(4260)$ to the $\bar D D_1(2420)+c.c.$ (see also the curve denoted by
``$Y$--loop''), which is an essential feature of the considered hadronic
molecular picture.
Although the cross section line shape plotted in Fig.~\ref{fig-4} is not
perfectly fitted, it can still clarify the role played by $Y(4260)$. The
dominance of the $\psi(4040)$ and $\psi(4160)$ near threshold actually leaves a
very limited space for the $Y(4260)$ which is consistent with the expectation
based on the hadronic molecule scenario for the $Y(4260)$. In other words, the
$Y(4260)$ does not have a large partial decay width in the $D^*\bar{D}^*$
channel.

We also tried a fit without including the $Y(4260)$, and found that the cross section at energies above the dip, which correspond to the region around the $\bar D D_1(2420)+c.c.$ threshold, cannot be well described. In fact, a negligibly small contribution from the $Y(4260)$ is consistent with the molecular picture.

From  Fig.~\ref{fig-4} we also see that the cross sections in the region of
$4.4\sim 4.6$ GeV can be well described by the interference between the
$\psi(4415)$ and the $Y(4260)$.
Despite this, we need to mention that the $S$-wave open thresholds of
$D^*\bar{D}_1(2420)+c.c.$ and $D^*\bar{D}_2(2460)+c.c.$ have not been taken into
account, and they could play a role in the region between 4.4 and 4.5~GeV. We
leave their contributions to be investigated more elaborately in future studies
when more data are available.

The transition of Fig.~\ref{fig-3} (b) has also access to the kinematics of the
so-called ``triangle singularity" (TS), which has been broadly investigated
recently in the
literature~\cite{Wu:2011yx,Wu:2012pg,Liu:2013vfa,Liu:2014spa,
Szczepaniak:2015eza,Liu:2015taa,Guo:2015umn,Liu:2015fea,Szczepaniak:2015hya,
Guo:2016bkl,Bayar:2016ftu,Yang:2016sws,Wang:2016dtb,Pilloni:2016obd,Gong:2016jzb}
(see e.g.
Refs.~\cite{Zhao:2017wey,Guo:2017jvc} for a recent review). For an appropriate input energy of the initial $e^+e^-$ annihilation, the TS condition
corresponds to that the internal particles can approach their on-shell
kinematics simultaneously and the interactions at all vertices can happen as
classical processes in space-time. With all the intermediate mesons fixed as
$D_1 \bar D\pi$ as in the figure and final states being $D^{*+}D^{*-}$, the
$e^+e^-$ c.m.
energy for producing a TS is at about 5.35~GeV which is far beyond the region of
Fig.~\ref{fig-4}. This situation is very different from the cases of $e^+e^-\to
\bar{D}D^*\pi$~\cite{Cleven:2013mka,Qin:2016spb} and
$J/\psi\pi\pi$~\cite{Wang:2013cya,Wang:2013hga}.

\begin{figure}[tb]
\centering
\scalebox{1}{\includegraphics{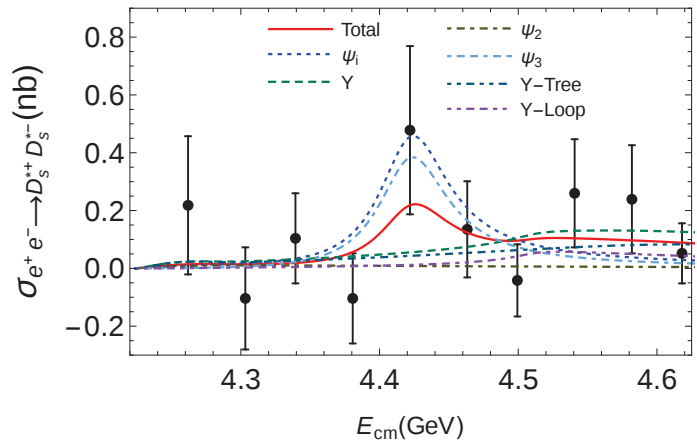}}\hspace{0.6cm}
\caption{The fitting results for the cross section of $e^+e^- \to D_s^{*+}D_s^{*-}$. The overall cross section is denoted by the solid line. The exclusive contributions from single states are also presented, i.e. $\psi(4160)$ (dot-dashed) and $\psi(4415)$ (dot-dashed-dashed), and $Y(4260)$ (dashed).
The inclusive contributions from $\psi_i$ is denoted by the dotted line. The data are from Ref.~\cite{Pakhlova:2010ek}.} \label{fig-5}
\end{figure}

As already mentioned the present experimental data for $e^+e^-\to
D_s^*\bar{D}_s^*$~\cite{Pakhlova:2010ek} do not allow a reliable determination
of the parameters in this channel. As shown in Fig.~\ref{fig-5}, with the
present fitted parameters only the $\psi(4415)$ can produce a resonance
structure in the cross section line shape. The exclusive contributions from
these three states are also presented. The theoretical curve shows a flattened
line shape near threshold which is different from that in $e^+e^-\to
D^*\bar{D}^*$. Although the $D_s^{*+}D_s^{*-}$ threshold, 4.22~GeV, is very
close to the mass of the $Y(4260)$, we do not see a near-threshold enhancement
due to the $Y(4260)$. We expect that the contribution of the $Y(4260)$ in this
process should be smaller than that in the $e^+e^-\to D^{*+}D^{*-}$ since the
intermediate kaon in Fig.~\ref{fig-3} (b) cannot go on shell, contrary to the
case of the pion. However, one notices that the poor data quality do not allow a
more quantitative restriction on the $Y(4260)$ contribution to this process. This situation is also reflected by the poor determination of the cutoff energy $\Lambda_K$ shown in Table~\ref{parameter-list}. A more precise measurement of the $e^+e^-\to D_s^*\bar{D}_s^*$ cross section line
shape is highly recommended.

\section{Summary}

In this work we have studied the cross section line shapes of the $e^+e^-\to
D^*\bar{D}^*$ and $D_s^*\bar{D}_s^*$ processes from thresholds to about 4.6~GeV.
This is the energy region that contributions from the $Y(4260)$  are of great
interest since information in addition to those in other processes about the
structure of this mysterious state can be extracted. Our study shows that the
cross sections of these two processes in this energy region are dominated by the
established charmonium states, i.e. $\psi(4040)$, $\psi(4160)$ and $\psi(4415)$,
while the contributions from the $Y(4260)$ as a $D\bar{D}_1(2420)+c.c.$ molecule
state turns out to be rather small in most of the energy region.
This result is consistent with the observation that the main open charm decay
channel of the $Y(4260)$ is $D\bar{D}^*\pi+c.c.$ which accounts for most of its
decay width. The partial decay width of the $Y(4260)\to D^*\bar D^*$ is obtained
to be $(11\pm8)$~MeV. The dip around 4.22~GeV in the $e^+e^-\to D^*\bar{D}^*$
cross section is due to the interference between the $Y(4260)$ and the
conventional charmonium states. The hadronic molecular feature of the $Y(4260)$
in our model shows up as a non-trivial structure at the $D\bar D_1(2420)$
threshold. The current data present a clear evidence for such a structure. Yet,
more precise data are necessary to make the conclusion more solid.
For the $e^+e^-\to D_s^*\bar{D}_s^*$ channel, the present experimental data from
Belle~\cite{Pakhlova:2010ek} have poor quality. However, although the data do
not allow any conclusion on the role played by charmonium states,
we do not expect sizeable contributions from the $Y(4260)$.
The future precise data from BESIII for these two channels will be able to
clarify the role played by the $Y(4260)$ and provide valuable insights into its
internal structure.

\section*{Acknowledgments}

Useful discussions with K.-T. Chao and C.-Z. Yuan are acknowledged. We are also
grateful to U.-G.~Mei{\ss}ner for a careful reading of the manuscript. This work
is supported, in part, by the National Natural Science Foundation of China
(NSFC) under Grant Nos. 11425525, 11521505 and 11647601, by DFG and NSFC through
funds provided to the Sino-German CRC 110 ``Symmetries and the Emergence of
Structure in QCD'' (NSFC Grant No.
11261130311), by the National Key Basic Research Program of China under Contract
No.~2015CB856700,  by the Thousand Talents Plan for Young Professionals, and by
the CAS Key Research Program of Frontier Sciences under Grant
No.~QYZDB-SSW-SYS013.


\begin{thebibliography}{99}

\bibitem{Aubert:2005rm}
  B.~Aubert {\it et al.} [BaBar Collaboration],
  %``Observation of a broad structure in the $\pi^+ \pi^- J/\psi$ mass spectrum around 4.26-GeV/c$^2$,''
  Phys.\ Rev.\ Lett.\  {\bf 95}, 142001 (2005)
%   doi:10.1103/PhysRevLett.95.142001
  [hep-ex/0506081].
  %%CITATION = doi:10.1103/PhysRevLett.95.142001;%%
  %688 citations counted in INSPIRE as of 09 Aug 2017


%\cite{Chen:2016qju}
\bibitem{Chen:2016qju}
  H.-X.~Chen, W.~Chen, X.~Liu and S.-L.~Zhu,
  %``The hidden-charm pentaquark and tetraquark states,''
  Phys.\ Rept.\  {\bf 639}, 1 (2016)
%   doi:10.1016/j.physrep.2016.05.004
  [arXiv:1601.02092 [hep-ph]].
  %%CITATION = doi:10.1016/j.physrep.2016.05.004;%%
  %178 citations counted in INSPIRE as of 09 Aug 2017


%\cite{Lebed:2016hpi}
\bibitem{Lebed:2016hpi}
  R.~F.~Lebed, R.~E.~Mitchell and E.~S.~Swanson,
  %``Heavy-Quark QCD Exotica,''
  Prog.\ Part.\ Nucl.\ Phys.\  {\bf 93}, 143 (2017)
%   doi:10.1016/j.ppnp.2016.11.003
  [arXiv:1610.04528 [hep-ph]].
  %%CITATION = doi:10.1016/j.ppnp.2016.11.003;%%
  %32 citations counted in INSPIRE as of 09 Aug 2017


%\cite{Esposito:2016noz}
\bibitem{Esposito:2016noz}
  A.~Esposito, A.~Pilloni and A.~D.~Polosa,
  %``Multiquark Resonances,''
  Phys.\ Rept.\  {\bf 668}, 1 (2016)
%   doi:10.1016/j.physrep.2016.11.002
  [arXiv:1611.07920 [hep-ph]].
  %%CITATION = doi:10.1016/j.physrep.2016.11.002;%%
  %28 citations counted in INSPIRE as of 09 Aug 2017


%\cite{Guo:2017jvc}
\bibitem{Guo:2017jvc}
  F.-K.~Guo, C.~Hanhart, U.-G.~Mei{\ss}ner, Q.~Wang, Q.~Zhao and B.~S.~Zou,
  %``Hadronic molecules,''
  arXiv:1705.00141 [hep-ph].
  %%CITATION = ARXIV:1705.00141;%%
  %17 citations counted in INSPIRE as of 09 Aug 2017


%\cite{Wang:2013kra}
\bibitem{Wang:2013kra}
  Q.~Wang, M.~Cleven, F.-K.~Guo, C.~Hanhart, U.-G.~Mei{\ss}ner, X.-G.~Wu and Q.~Zhao,
  %``Y(4260): hadronic molecule versus hadro-charmonium interpretation,''
  Phys.\ Rev.\ D {\bf 89},  034001 (2014)
%   doi:10.1103/PhysRevD.89.034001
  [arXiv:1309.4303 [hep-ph]].
  %%CITATION = doi:10.1103/PhysRevD.89.034001;%%
  %30 citations counted in INSPIRE as of 09 Aug 2017


%\cite{Cleven:2013mka}
\bibitem{Cleven:2013mka}
  M.~Cleven, Q.~Wang, F.-K.~Guo, C.~Hanhart, U.-G.~Mei{\ss}ner and Q.~Zhao,
  %``$Y(4260)$ as the first $S$-wave open charm vector molecular state?,''
  Phys.\ Rev.\ D {\bf 90},  074039 (2014)
%   doi:10.1103/PhysRevD.90.074039
  [arXiv:1310.2190 [hep-ph]].
  %%CITATION = doi:10.1103/PhysRevD.90.074039;%%
  %38 citations counted in INSPIRE as of 09 Aug 2017


%\cite{Qin:2016spb}
\bibitem{Qin:2016spb}
  W.~Qin, S.-R.~Xue and Q.~Zhao,
  %``Production of $Y(4260)$ as a hadronic molecule state of $\bar{D}D_1 +c.c.$ in $e^+e^-$ annihilations,''
  Phys.\ Rev.\ D {\bf 94},  054035 (2016)
%   doi:10.1103/PhysRevD.94.054035
  [arXiv:1605.02407 [hep-ph]].
  %%CITATION = doi:10.1103/PhysRevD.94.054035;%%
  %8 citations counted in INSPIRE as of 09 Aug 2017


%\cite{Ablikim:2013mio}
\bibitem{Ablikim:2013mio}
  M.~Ablikim {\it et al.} [BESIII Collaboration],
  %``Observation of a Charged Charmoniumlike Structure in $e^+e^-$ ''
  Phys.\ Rev.\ Lett.\  {\bf 110}, 252001 (2013)
%   doi:10.1103/PhysRevLett.110.252001
  [arXiv:1303.5949 [hep-ex]].
  %%CITATION = doi:10.1103/PhysRevLett.110.252001;%%
  %501 citations counted in INSPIRE as of 09 Aug 2017


%\cite{Ablikim:2013wzq}
\bibitem{Ablikim:2013wzq}
  M.~Ablikim {\it et al.} [BESIII Collaboration],
  %``Observation of a Charged Charmoniumlike Structure $Z_c$(4020) and Search for the $Z_c$(3900) in $e^+e^- \to \pi^+\pi^-h_c$,''
  Phys.\ Rev.\ Lett.\  {\bf 111},  242001 (2013)
%   doi:10.1103/PhysRevLett.111.242001
  [arXiv:1309.1896 [hep-ex]].
  %%CITATION = doi:10.1103/PhysRevLett.111.242001;%%
  %239 citations counted in INSPIRE as of 09 Aug 2017


%\cite{Ablikim:2013xfr}
\bibitem{Ablikim:2013xfr}
  M.~Ablikim {\it et al.} [BESIII Collaboration],
  %``Observation of a charged $(D\bar{D}^{*})^\pm$ mass peak in $e^{+}e^{-} \to \pi D\bar{D}^{*}$ at $\sqrt{s} =$ 4.26 GeV,''
  Phys.\ Rev.\ Lett.\  {\bf 112},  022001 (2014)
%   doi:10.1103/PhysRevLett.112.022001
  [arXiv:1310.1163 [hep-ex]].
  %%CITATION = doi:10.1103/PhysRevLett.112.022001;%%
  %189 citations counted in INSPIRE as of 09 Aug 2017


%\cite{Ablikim:2014dxl}
\bibitem{Ablikim:2014dxl}
  M.~Ablikim {\it et al.} [BESIII Collaboration],
  %``Observation of $e^+e^- \to \pi^0\pi^0h_c$ and a Neutral Charmoniumlike Structure $Z_c(4020)^0$,''
  Phys.\ Rev.\ Lett.\  {\bf 113},  212002 (2014)
%   doi:10.1103/PhysRevLett.113.212002
  [arXiv:1409.6577 [hep-ex]].
  %%CITATION = doi:10.1103/PhysRevLett.113.212002;%%
  %77 citations counted in INSPIRE as of 09 Aug 2017


%\cite{Ablikim:2015tbp}
\bibitem{Ablikim:2015tbp}
  M.~Ablikim {\it et al.} [BESIII Collaboration],
  %``Observation of $Z_c(3900)^{0}$ in $e^+e^-\to\pi^0\pi^0 J/\psi$,''
  Phys.\ Rev.\ Lett.\  {\bf 115},  112003 (2015)
%   doi:10.1103/PhysRevLett.115.112003
  [arXiv:1506.06018 [hep-ex]].
  %%CITATION = doi:10.1103/PhysRevLett.115.112003;%%
  %55 citations counted in INSPIRE as of 09 Aug 2017


%\cite{Ablikim:2015vvn}
\bibitem{Ablikim:2015vvn}
  M.~Ablikim {\it et al.} [BESIII Collaboration],
  %``Observation of a neutral charmoniumlike state $Z_c(4025)^0$ in $e^{+} e^{-} \to (D^{*} \bar{D}^{*})^{0} \pi^0$,''
  Phys.\ Rev.\ Lett.\  {\bf 115},  182002 (2015)
%   doi:10.1103/PhysRevLett.115.182002
  [arXiv:1507.02404 [hep-ex]].
  %%CITATION = doi:10.1103/PhysRevLett.115.182002;%%
  %41 citations counted in INSPIRE as of 09 Aug 2017



%\cite{Gao:2017sqa}
\bibitem{Gao:2017sqa}
  X.~Y.~Gao, C.~P.~Shen and C.~Z.~Yuan,
  %``Resonant parameters of the $Y(4220)$,''
  Phys.\ Rev.\ D {\bf 95}, no. 9, 092007 (2017)
  doi:10.1103/PhysRevD.95.092007
  [arXiv:1703.10351 [hep-ex]].

%\cite{Abe:2006fj}
\bibitem{Abe:2006fj}
  K.~Abe {\it et al.} [Belle Collaboration],
  %``Measurement of the near-threshold e+ e- ---> D(*)+- D(*)-+ cross section using initial-state radiation,''
  Phys.\ Rev.\ Lett.\  {\bf 98}, 092001 (2007)
%   doi:10.1103/PhysRevLett.98.092001
  [hep-ex/0608018].
  %%CITATION = doi:10.1103/PhysRevLett.98.092001;%%
  %144 citations counted in INSPIRE as of 09 Aug 2017


%\cite{Pakhlova:2010ek}
\bibitem{Pakhlova:2010ek}
  G.~Pakhlova {\it et al.} [Belle Collaboration],
  %``Measurement of $e^+e^-\to D_s^{(*)+} D_s^{(*)-}$ cross sections near threshold using initial-state radiation,''
  Phys.\ Rev.\ D {\bf 83}, 011101 (2011)
%   doi:10.1103/PhysRevD.83.011101
  [arXiv:1011.4397 [hep-ex]].
  %%CITATION = doi:10.1103/PhysRevD.83.011101;%%
  %19 citations counted in INSPIRE as of 09 Aug 2017

\bibitem{Du:2016qcr}
  M.-L.~Du, U.-G.~Mei{\ss}ner and Q.~Wang,
  %``$P$-wave coupled channel effects in electron-positron annihilation,''
  Phys.\ Rev.\ D {\bf 94}, 096006 (2016)
%   doi:10.1103/PhysRevD.94.096006
  [arXiv:1608.02537 [hep-ph]].

%\cite{Wang:2013cya}
\bibitem{Wang:2013cya}
  Q.~Wang, C.~Hanhart and Q.~Zhao,
  %``Decoding the riddle of $Y(4260)$ and $Z_c(3900)$,''
  Phys.\ Rev.\ Lett.\  {\bf 111},  132003 (2013)
%   doi:10.1103/PhysRevLett.111.132003
  [arXiv:1303.6355 [hep-ph]].
  %%CITATION = doi:10.1103/PhysRevLett.111.132003;%%
  %169 citations counted in INSPIRE as of 09 Aug 2017

%\cite{Guo:2013nza}
\bibitem{Guo:2013nza}
  F.-K.~Guo, C.~Hanhart, U.-G.~Mei{\ss}ner, Q.~Wang and Q.~Zhao,
  %``Production of the X(3872) in charmonia radiative decays,''
  Phys.\ Lett.\ B {\bf 725}, 127 (2013)
%   doi:10.1016/j.physletb.2013.06.053
  [arXiv:1306.3096 [hep-ph]].
  %%CITATION = doi:10.1016/j.physletb.2013.06.053;%%
  %61 citations counted in INSPIRE as of 09 Aug 2017

%\cite{Cleven:2016qbn}
\bibitem{Cleven:2016qbn}
  M.~Cleven and Q.~Zhao,
  %``Cross section line shape of $e^+e^-\to\chi_{c0}\omega$ around the $Y(4260)$ mass region,''
  Phys.\ Lett.\ B {\bf 768}, 52 (2017)
%   doi:10.1016/j.physletb.2017.02.041
  [arXiv:1611.04408 [hep-ph]].
  %%CITATION = doi:10.1016/j.physletb.2017.02.041;%%
  %2 citations counted in INSPIRE as of 09 Aug 2017

\bibitem{Guo:2009wr}
  F.-K.~Guo, C.~Hanhart and U.-G.~Mei{\ss}ner,
  %``On the extraction of the light quark mass ratio from the decays psi-prime ---> J/psi pi0 (eta),''
  Phys.\ Rev.\ Lett.\  {\bf 103}, 082003 (2009)
  Erratum: [Phys.\ Rev.\ Lett.\  {\bf 104}, 109901 (2010)]
%   doi:10.1103/PhysRevLett.103.082003
  [arXiv:0907.0521 [hep-ph]].

%\cite{Guo:2010ak}
\bibitem{Guo:2010ak}
  F.-K.~Guo, C.~Hanhart, G.~Li, U.-G.~Mei{\ss}ner and Q.~Zhao,
  %``Effect of charmed meson loops on charmonium transitions,''
  Phys.\ Rev.\ D {\bf 83}, 034013 (2011)
%   doi:10.1103/PhysRevD.83.034013
  [arXiv:1008.3632 [hep-ph]].
  %%CITATION = doi:10.1103/PhysRevD.83.034013;%%
  %76 citations counted in INSPIRE as of 09 Aug 2017

%\cite{Li:2013yka}
\bibitem{Li:2013yka}
  X.~Li and M.~B.~Voloshin,
  %``Suppression of the $S$-wave production of $(3/2)^+$ + $(1/2)^-$ heavy meson pairs in $e^+e^-$ annihilation,''
  Phys.\ Rev.\ D {\bf 88}, 034012 (2013)
%   doi:10.1103/PhysRevD.88.034012
  [arXiv:1307.1072 [hep-ph]].

\bibitem{Margaryan:2013tta}
  A.~Margaryan and R.~P.~Springer,
  %``Using the decay ψ(4160)→X(3872) γ to probe the molecular content of the X(3872),''
  Phys.\ Rev.\ D {\bf 88}, 014017 (2013)
%   doi:10.1103/PhysRevD.88.014017
  [arXiv:1304.8101 [hep-ph]].


\bibitem{Cleven:2011gp}
  M.~Cleven, F.-K.~Guo, C.~Hanhart and U.-G.~Mei{\ss}ner,
  %``Bound state nature of the exotic $Z_b$ states,''
  Eur.\ Phys.\ J.\ A {\bf 47}, 120 (2011)
%   doi:10.1140/epja/i2011-11120-6
  [arXiv:1107.0254 [hep-ph]].

%\cite{Cao:2016xqo}
\bibitem{Cao:2016xqo}
  Z.~Cao, M.~Cleven, Q.~Wang and Q.~Zhao,
  %``Open charm contributions to the E1 transitions of $\psi (3686)$ and $\psi (3770)\rightarrow \gamma \chi _{cJ}$,''
  Eur.\ Phys.\ J.\ C {\bf 76},  601 (2016)
%   doi:10.1140/epjc/s10052-016-4448-3
  [arXiv:1608.07947 [hep-ph]].
  %%CITATION = doi:10.1140/epjc/s10052-016-4448-3;%%
  %5 citations counted in INSPIRE as of 09 Aug 2017


%\cite{Olive:2016xmw}
\bibitem{Olive:2016xmw}
  C.~Patrignani {\it et al.} [Particle Data Group],
  %``Review of Particle Physics,''
  Chin.\ Phys.\ C {\bf 40},  100001 (2016).
%   doi:10.1088/1674-1137/40/10/100001
  %%CITATION = doi:10.1088/1674-1137/40/10/100001;%%
  %1318 citations counted in INSPIRE as of 09 Aug 2017


%\cite{Wu:2011yx}
\bibitem{Wu:2011yx}
  J.-J.~Wu, X.-H.~Liu, Q.~Zhao and B.-S.~Zou,
  %``The Puzzle of anomalously large isospin violations in $\eta(1405/1475)\to 3\pi$,''
  Phys.\ Rev.\ Lett.\  {\bf 108}, 081803 (2012)
%   doi:10.1103/PhysRevLett.108.081803
  [arXiv:1108.3772 [hep-ph]].
  %%CITATION = doi:10.1103/PhysRevLett.108.081803;%%
  %75 citations counted in INSPIRE as of 09 Aug 2017


%\cite{Wu:2012pg}
\bibitem{Wu:2012pg}
  X.-G.~Wu, J.-J.~Wu, Q.~Zhao and B.-S.~Zou,
  %``Understanding the property of $\eta(1405/1475)$ in the $J/\psi$ radiative decay,''
  Phys.\ Rev.\ D {\bf 87},  014023 (2013)
%   doi:10.1103/PhysRevD.87.014023
  [arXiv:1211.2148 [hep-ph]].
  %%CITATION = doi:10.1103/PhysRevD.87.014023;%%
  %29 citations counted in INSPIRE as of 09 Aug 2017

%\cite{Liu:2013vfa}
\bibitem{Liu:2013vfa}
  X.-H.~Liu and G.~Li,
  %``Exploring the threshold behavior and implications on the nature of Y(4260) and Zc(3900),''
  Phys.\ Rev.\ D {\bf 88}, 014013 (2013)
%   doi:10.1103/PhysRevD.88.014013
  [arXiv:1306.1384 [hep-ph]].
  %%CITATION = doi:10.1103/PhysRevD.88.014013;%%
  %55 citations counted in INSPIRE as of 09 Aug 2017

\bibitem{Liu:2014spa}
  X.~H.~Liu,
  %``Influence of threshold effects induced by charmed meson rescattering,''
  Phys.\ Rev.\ D {\bf 90}, 074004 (2014)
%   doi:10.1103/PhysRevD.90.074004
  [arXiv:1403.2818 [hep-ph]].

\bibitem{Szczepaniak:2015eza}
  A.~P.~Szczepaniak,
  %``Triangle Singularities and XYZ Quarkonium Peaks,''
  Phys.\ Lett.\ B {\bf 747}, 410 (2015)
%   doi:10.1016/j.physletb.2015.06.029
  [arXiv:1501.01691 [hep-ph]].

%\cite{Liu:2015taa}
\bibitem{Liu:2015taa}
  X.-H.~Liu, M.~Oka and Q.~Zhao,
  %``Searching for observable effects induced by anomalous triangle singularities,''
  Phys.\ Lett.\ B {\bf 753}, 297 (2016)
%   doi:10.1016/j.physletb.2015.12.027
  [arXiv:1507.01674 [hep-ph]].
  %%CITATION = doi:10.1016/j.physletb.2015.12.027;%%
  %35 citations counted in INSPIRE as of 09 Aug 2017

%\cite{Guo:2015umn}
\bibitem{Guo:2015umn}
  F.-K.~Guo, U.-G.~Mei{\ss}ner, W.~Wang and Z.~Yang,
  %``How to reveal the exotic nature of the P$_c$(4450),''
  Phys.\ Rev.\ D {\bf 92},  071502 (2015)
%   doi:10.1103/PhysRevD.92.071502
  [arXiv:1507.04950 [hep-ph]].
  %%CITATION = doi:10.1103/PhysRevD.92.071502;%%
  %118 citations counted in INSPIRE as of 09 Aug 2017


%\cite{Liu:2015fea}
\bibitem{Liu:2015fea}
  X.-H.~Liu, Q.~Wang and Q.~Zhao,
  %``Understanding the newly observed heavy pentaquark candidates,''
  Phys.\ Lett.\ B {\bf 757}, 231 (2016)
%   doi:10.1016/j.physletb.2016.03.089
  [arXiv:1507.05359 [hep-ph]].
  %%CITATION = doi:10.1016/j.physletb.2016.03.089;%%
  %101 citations counted in INSPIRE as of 09 Aug 2017

\bibitem{Szczepaniak:2015hya}
  A.~P.~Szczepaniak,
  %``Dalitz plot distributions in presence of triangle singularities,''
  Phys.\ Lett.\ B {\bf 757}, 61 (2016)
%   doi:10.1016/j.physletb.2016.03.064
  [arXiv:1510.01789 [hep-ph]].

\bibitem{Guo:2016bkl}
  F.-K.~Guo, U.-G.~Mei{\ss}ner, J.~Nieves and Z.~Yang,
  %``Remarks on the $P_c$ structures and triangle singularities,''
  Eur.\ Phys.\ J.\ A {\bf 52}, 318 (2016)
%   doi:10.1140/epja/i2016-16318-4
  [arXiv:1605.05113 [hep-ph]].

%\cite{Bayar:2016ftu}
\bibitem{Bayar:2016ftu}
  M.~Bayar, F.~Aceti, F.-K.~Guo and E.~Oset,
  %``A Discussion on Triangle Singularities in the $\Lambda_b \to J/\psi K^{-} p$ Reaction,''
  Phys.\ Rev.\ D {\bf 94},  074039 (2016)
%   doi:10.1103/PhysRevD.94.074039
  [arXiv:1609.04133 [hep-ph]].
  %%CITATION = doi:10.1103/PhysRevD.94.074039;%%
  %13 citations counted in INSPIRE as of 09 Aug 2017


%\cite{Yang:2016sws}
\bibitem{Yang:2016sws}
  Z.~Yang, Q.~Wang and U.-G.~Mei{\ss}ner,
  %``Where does the X(5568) structure come from?,''
  Phys.\ Lett.\ B {\bf 767}, 470 (2017)
%   doi:10.1016/j.physletb.2017.01.023
  [arXiv:1609.08807 [hep-ph]].
  %%CITATION = doi:10.1016/j.physletb.2017.01.023;%%
  %5 citations counted in INSPIRE as of 09 Aug 2017

%\cite{Wang:2016dtb}
\bibitem{Wang:2016dtb}
  E.~Wang, J.~J.~Xie, W.~H.~Liang, F.-K.~Guo and E.~Oset,
  %``Role of a triangle singularity in the $\gamma p\rightarrow K^+ \Lambda(1405)$ reaction,''
  Phys.\ Rev.\ C {\bf 95},  015205 (2017)
%   doi:10.1103/PhysRevC.95.015205
  [arXiv:1610.07117 [hep-ph]].
  %%CITATION = doi:10.1103/PhysRevC.95.015205;%%
  %9 citations counted in INSPIRE as of 09 Aug 2017

\bibitem{Pilloni:2016obd}
  A.~Pilloni {\it et al.} [JPAC Collaboration],
  %``Amplitude analysis and the nature of the Z$_c$(3900),''
  Phys.\ Lett.\ B {\bf 772}, 200 (2017)
%   doi:10.1016/j.physletb.2017.06.030
  [arXiv:1612.06490 [hep-ph]].

\bibitem{Gong:2016jzb}
  Q.~R.~Gong, J.~L.~Pang, Y.~F.~Wang and H.~Q.~Zheng,
  %``The $Z_c(3900)$ peak does not come from the "triangle singularity",''
  arXiv:1612.08159 [hep-ph].

%\cite{Zhao:2017wey}
\bibitem{Zhao:2017wey}
  Q.~Zhao,
  %``Threshold and Kinematic Effects for the Exotic Resonance-Like Structures,''
  JPS Conf.\ Proc.\  {\bf 13}, 010008 (2017).
%   doi:10.7566/JPSCP.13.010008
  %%CITATION = doi:10.7566/JPSCP.13.010008;%%
  %1 citations counted in INSPIRE as of 09 Aug 2017


%\cite{Wang:2013hga}
\bibitem{Wang:2013hga}
  Q.~Wang, C.~Hanhart and Q.~Zhao,
  %``Systematic study of the singularity mechanism in heavy quarkonium decays,''
  Phys.\ Lett.\ B {\bf 725},  106 (2013)
%   doi:10.1016/j.physletb.2013.06.049
  [arXiv:1305.1997 [hep-ph]].





\end{thebibliography}
\end{document}